\documentclass[twocolumn,showpacs,aps,prl,superscriptaddress]{revtex4}
\usepackage{graphicx,psfrag,amsmath,amssymb,float}
\usepackage{color}
\input{epsf}
\newcommand{\bc}{\begin{center}}
\newcommand{\ec}{\end{center}}
\newcommand{\be}{\begin{equation}}
\newcommand{\ee}{\end{equation}}
\newcommand{\bea}{\begin{eqnarray}}
\newcommand{\eea}{\end{eqnarray}}

\def\1.2{\frac{1}{2}}

\begin{document}
\title{Impurity Entanglement Entropy and the Kondo Screening Cloud}
\author{Erik S. S{\o}rensen}
\affiliation{Department of Physics and Astronomy, McMaster
University, Hamilton, ON, L8S 4M1 Canada}
\author{Ming-Shyang Chang}
\affiliation{Department of Physics \& Astronomy, University of
British Columbia, Vancouver, B.C., Canada, V6T 1Z1}
\author{Nicolas Laflorencie}
\affiliation{Department of Physics \& Astronomy, University of
British Columbia, Vancouver, B.C., Canada, V6T 1Z1}
\author{Ian Affleck}
\affiliation{Department of Physics \& Astronomy, University of
British Columbia, Vancouver, B.C., Canada, V6T 1Z1}
\date{\today}
\begin{abstract}
The screening of an impurity spin by conduction electrons is
associated with the formation of a large Kondo screening cloud, of
size $\xi_K$. We study the quantum entanglement between a
region of size $r$ surrounding the impurity and the rest
of the sample, (of total size ${\mathcal R}$) using Density Matrix
Renormalization Group and analytic methods.  The corresponding
``impurity entanglement entropy'', $S_{imp}$, is shown to be a universal
scaling function of $r/\xi_K$ and $r/{\mathcal R}$. We calculate this universal
function using Fermi liquid theory in the regime $\xi_K\ll r$.
\end{abstract}
\pacs{03.67.Mn,75.30.Hx,75.10.Pq}
\maketitle

Quantum entanglement of many body wave-functions is receiving much attention
of late due to its quantum information (QI) theory applications \cite{Bennett00}, its
importance to computational algorithms \cite{Daley} and
 the new perspective that it 
offers on critical behavior of condensed matter systems \cite{Vidal03,Cardy04}. 
Very recently, entanglement has been investigated in some quantum impurity models \cite{Levine}. 
The Kondo Hamiltonian \cite{Hewson}, a famous quantum impurity model, was originally proposed 
to represent a single impurity spin interacting with conduction electrons in a metal. The Hamiltonian is:
\begin{equation}
H=\int d^{3}r[\psi ^{\dagger }(-\nabla ^{2}/2m)\psi +J _{K}\delta^3(r)\psi
^{\dagger }(\vec{\sigma}/2)\psi \cdot \vec{S}],  \label{H3D}
\end{equation}
where  $\psi (\vec{r})$ is the electron annihilation operator (with
spin-index suppressed), the $S^{a}$ are $S=1/2$ spin operators and 
the Kondo coupling, $J_K>0$. Due to the $\delta$-function 
in Eq. (\ref{H3D}), we may expand $\psi (\vec r)$ in spherical harmonics and only the s-wave couples to 
the impurity, yielding an effective one dimensional (1$D$) model.  From spin-charge separation 
in one dimension, it can be seen that 
the universal low energy long distance physics of the Kondo model  also 
occurs in the $J_1-J_2$ family of gapless antiferromagnetic $S=1/2$ Heisenberg spin chain models for
$J_2\leq J_2^c\simeq0.2412$,
with the coupling of the end spin to its neighbors, $J_K^\prime$,
weaker than the other couplings \cite{Eggert}:
\begin{eqnarray}
H=&&
J_{K}^{\prime }\left(\vec{S}_{1}\cdot \vec{S}_{2}+J_{2}\vec{S}_{1}\cdot \vec{S}_{3}\right)+\nonumber\\
&&\sum_{j=2}^{{\mathcal R}-1}\vec{S}_{j}\cdot \vec{S}_{j+1}+
J_2\sum_{j=2}^{{\mathcal R}-2}\vec{S}_{j}\cdot \vec{S}_{j+2}.\label{spinch}
\end{eqnarray}
The Kondo
coupling, $J_{K}$ (or $J_K^\prime$), is found to grow large at low energies
under renormalization and the ground state is a highly non-trivial many body
state in which the impurity spin is sometimes heuristically thought of as
being screened by a conduction electron in a large orbital of size
\begin{figure}[th]
\begin{center}
\includegraphics[width=\columnwidth,clip]{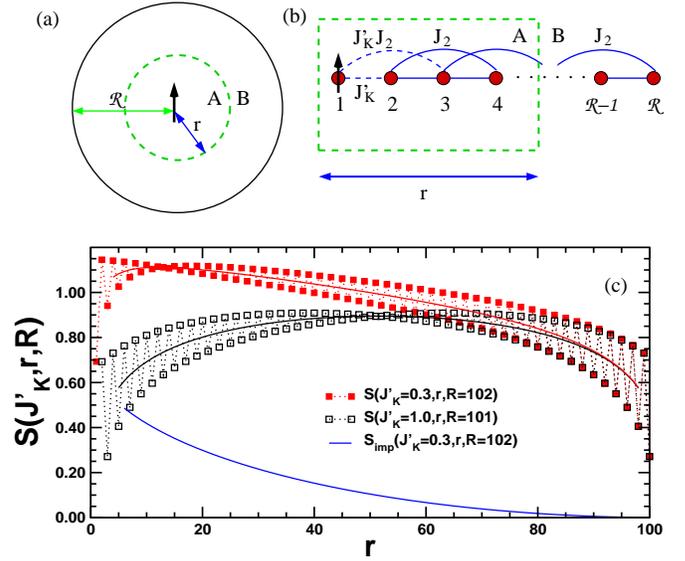}
\end{center}
\caption{
(a) Schematic picture of the two regions A,B around a Kondo Impurity.
(b) 1$D$ analogue of the Kondo problem, with 
${\mathcal{R}}$ interacting qubits along an antiferromagnetic chain. 
(c) Total entanglement entropy, $S(J'_K,r,{\mathcal R})$ for a 102 site spin chain at $J^c_2$,
with a $J'_K=0.3$ Kondo impurity along with $S(J'_K=1,r-1,{\mathcal R}-1)$ (no impurity). 
Uniform parts (solid lines) and the resulting $S_{imp}$.}
\label{fig:setup}
\end{figure}
\begin{equation}
\xi _{K}=v/T_{K}.  \label{xi}
\end{equation}
Here $v$ is the velocity of the low energy excitations (fermions or spin-waves) 
and $T_{K}$ is the energy scale at
which the renormalized Kondo coupling becomes large. For weak 
bare coupling, $\xi_K\propto e^{\alpha /J_K}$ where $\alpha$ is a constant.
Many physical observables were shown to vary on the characteristic 
length scale, $\xi_K$
 \cite{Sorensen, Barzykin, Simon}. 

\begin{figure}[th]
\psfrag{J}{$\mathcal J$}
\begin{center}
\includegraphics[width=0.95\columnwidth,clip]{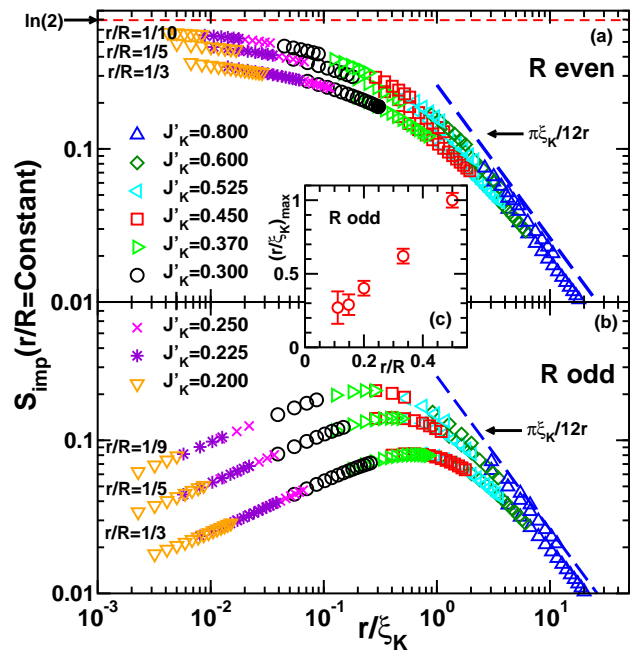}
\end{center}
\caption{Universal scaling plot of $S_{\mathrm{imp}}$ for fixed $r/{\mathcal R}$,
(a) for $\mathcal{R}$ {\it even}, (b) for $\mathcal{R}$ {\it odd}.
DMRG results for the $J_{1}-J_{2}$ chain at $J^c_2$
for various couplings $J_{K}^{\prime }$. 
The lines marked $ \pi \xi_K/(12r)$ are the FLT
prediction: Eq.~(\ref{eq:FLT_FSS}). (c): the location of the maximum, $(r/\xi_K)_{\rm max}$,
of $S_{imp}$ for odd $\mathcal{R}$, plotted versus $r/\mathcal{R}$. 
} \label{fig:Scaling1}
\end{figure}

The purpose of this paper is to investigate the interplay of the 
Kondo screening cloud and quantum entanglement occurring in the models Eqs. (\ref{H3D}-\ref{spinch}). 
In a QI context we could think of the three dimensional (3$D$) fermionic model, Eq.~(\ref{H3D}), as representing 
a single qubit coupled to a dissipative environment.  Alternatively, 
the 1$D$ spin model, Eq.~(\ref{spinch}), corresponds to a chain of qubits with 
the end qubit more weakly coupled than the rest.
$S=1/2$ chains are receiving considerable attention from the QI community
in the context of perfect state transfer~\cite{Transfer}.
In the 3D fermionic model, we imagine separating space into a
spherical region, $A$, of radius $r$ surrounding the impurity spin
and the region $B$, exterior to the sphere as in 
Fig.~\ref{fig:setup}(a). 
Tracing
out the region $B$ from the (pure) ground state density matrix, we
obtain the reduced density matrix $\rho (r)$ for $A$. The
entanglement entropy is defined as its Neumann entropy
$S=-\hbox{tr}[\rho (r)\ln \rho (r)]$ \cite{Bennett96}.  
In the 1$D$ spin
chain version, we consider the entanglement of the 
first $r$ spins with the remainder of the chain
as in Fig.~\ref{fig:setup}(b). 
For the 1$D$ spin chain at $J_2^c$ we show the resulting entanglement entropy in Fig.~\ref{fig:setup}(c)
for $J_K^\prime =.3$ and with no impurity for $J'_K=1$. 
Note that, 
in both cases, there is a strong even-odd alternation of $S$ with $r$. 
This effect  
was  studied in \cite{Nicolas} for the case of uniform 
coupling, $J_K^\prime =1$. In the present paper we focus on the 
uniform part, which can be defined at $r\gg 1$ by:
\begin{equation} S(r,{\mathcal R})=S_U(J'_K,r,{\mathcal{R}})+(-1)^rS_A(J'_K,r,{\mathcal{R}}),\end{equation}
where $S_U(J'_K,r,{\mathcal{R}})$ and $S_A(J'_K,r,{\mathcal{R}})$ are slowly varying functions
or $r$. There is also a strong alternation with the parity of
${\mathcal R}$ and we study both ${\mathcal R}$ even and odd. 
The uniform part consists of a bulk part, 
$
S_{U0}(r,{\mathcal{R}})\approx (1/6)\ln [{\mathcal{R}} \sin (\pi r/{\mathcal{R}})]+\hbox{constant},
  $
independent of $J_K^\prime$ \cite{Wilczek94,Cardy04,Nicolas}
and an impurity term that we define precisely as: 
\begin{equation}
S_{imp}(J'_K,r,{\mathcal{R}})\equiv S_U(J'_K,r,{\mathcal{R}})-S_{U}(1,r-1,{\mathcal{R}}-1).\label{Simpdef}
  \end{equation}
Here we have subtracted $S_U$ when the impurity is absent.
See Fig.~\ref{fig:setup}(c).
There is also an impurity term in $S_A$ but we do not address this here.
We focus here on calculating $S_{imp}$ for the spin chain model but the same universal scaling 
function also describes the additional entanglement entropy due to the impurity in the 
3D Kondo model \cite{SCLA}. 
We find that $S_{imp}(J'_K,r,{\mathcal{R}})$ 
appears to be a scaling function, depending only on the ratios of characteristic lengths
when $r$, $\xi_K\gg 1$:
\begin{equation} 
S_{imp}(J^{\prime}_K,r,{\mathcal{R}})=S_{imp}(r/\xi_K,r/{\mathcal{R}}).\label{eq:scaling}
\end{equation}
Note that there are actually two different scaling functions 
for the cases of ${\mathcal R}$ even or odd. 
We have verified this scaling form over a wide range of parameters using the
Density Matrix Renormalization Group (DMRG) keeping from 256 to 1024 
states (Fig.~\ref{fig:Scaling1}).
In the limit $\xi _{K}\ll r$ we have calculated the universal scaling
function explicitly using conformal field theory (CFT) methods based on
Nozi\`{e}res' local Fermi liquid theory (FLT) \cite{Nozieres, IanCFT}.

{\it Numerical Results.}
We have determined $S_{imp}$ for a range of values of $J'_{K}$, $r$ and $%
\mathcal{R}$ using DMRG on the spin chain model, Eq. (\ref{spinch}).  
In order to avoid well-known logarithmic corrections due to
a marginally irrelevant (bulk) operator \cite{Cardy86} we 
tune $J_2$ to the critical point $J^c_{2}\approx .2412$,
where the marginal coupling constant
vanishes \cite{Eggert}. We then extract $S_U$ from the DMRG data
with high precision using a 7-point formula. 
In Fig.~\ref{fig:Scaling1} we show a scaling plot of $S_{imp}$ for several fixed $r/{\mathcal R}$
allowing for a determination of $\xi_K(J'_K)$ and validating the scaling form, Eq.~(\ref{eq:scaling}).

\begin{figure}[!th]
\begin{center}
\includegraphics[width=\columnwidth,clip]{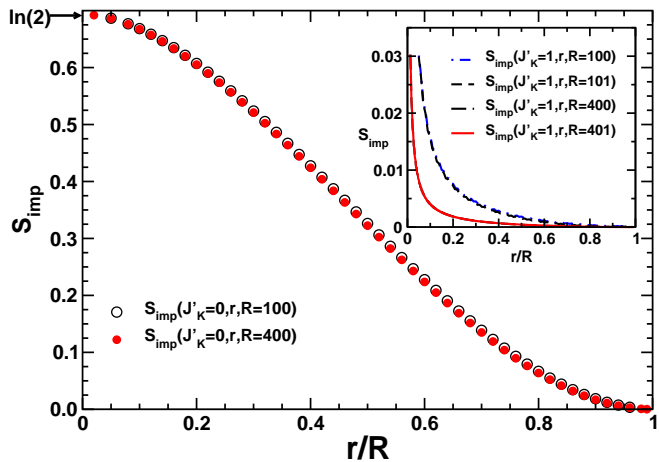}
\end{center}
\caption{
DMRG results using spin-inversion for $S_{\mathrm{imp}}(0,r,\mathcal{R})$  for $\mathcal{R}$ even. 
Inset: DMRG results for $S_{\mathrm{imp}}(1,r,\mathcal{R})$  for $\mathcal{R}$ odd and even. Data for $\mathcal{R}=100,101$ 
and $\mathcal{R}=400,401$ are indistinguishable. 
$S_{\mathrm{imp}}(1,r,\mathcal{R})$  appears to vanish in the scaling limit $r\to\infty$ with $r/\mathcal{R}$
held fixed.
} \label{fig:sfp}
\end{figure}

For odd $\mathcal{R}$, $S(0,r,{\mathcal R})-S(1,r-1,{\mathcal R}-1)$ is 
exactly zero, since the decoupled impurity at $j=1$ has no 
entanglement with the rest of the chain in the $S=1/2$ 
ground state in which it is fully polarized.  This implies 
that $S_{imp}(0,r,{\mathcal R})=0$ for ${\mathcal R}$ odd.
On the other hand, for ${\mathcal R}$ even, 
$S_{imp}(0,r,{\mathcal R})$ is not zero, as shown in Fig.~\ref{fig:sfp}. 
Thus the presence of the impurity spin, even when there is no term in the 
Hamiltonian coupling it to the rest of the system, alters $S$.  The reason 
this is possible is because, with ${\mathcal R}$ even 
 and $J_K^{\prime}=0$, the ground state 
is 4-fold degenerate, corresponding to the impurity spin and the rest of the chain 
independently 
having $S^z=\pm 1/2$.  The ground state that we consider in this case is the 
spin singlet 
state.  This is necessary to connect to the limit $J_K^{\prime}\to 0^+$. 
$S(0,r,\mathcal{R})$ is then clearly different from $S(1,r-1,\mathcal{R}-1)$
with a resulting non-zero $S_{imp}(0,r,\mathcal{R})$ for $\mathcal{R}$ even~\cite{SCLA}.
We expect that, in the limit $r/{\mathcal R}\to 0$, 
$S_{imp}(r/\xi_K)$ becomes the same for even and odd $R$, since 
adding one more site to an infinite chain should not have 
any effect on the entanglement entropy.
Our data suggests that $\lim_{r/\xi_K\to 0}\lim_{r/R\to 0}S_{imp}(r/\xi_K,r/{\mathcal R})=\ln 2$, 
for either parity of ${\mathcal R}$.
We note that the 
presence of a term in the entanglement entropy at a 
conformally invariant boundary fixed point, corresponding 
to the zero temperature thermodynamic impurity entropy, $\ln g$ 
was pointed out in Ref. \cite{Cardy04}.  For the single 
channel Kondo model $\ln g = \ln 2$ at the weak 
coupling fixed point where the impurity is unscreened, and 
$\ln g = 0$ at the strong coupling fixed point where 
it is screened. The thermodynamic
impurity entropy decreases monotonically from $\ln 2$ to $0$ with decreasing $T/T_K$.
The impurity entanglement entropy behaves the same way with increasing $r/\xi_K$.

\begin{figure}[th]
\begin{center}
\includegraphics[width=\columnwidth,clip]{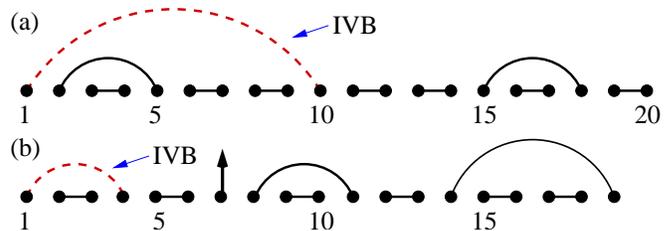}
\end{center}
\caption {
Typical valence bond configurations for $\mathcal{R}$ even (a) and
$\mathcal{R}$ odd (b) where the unpaired spin is on the $7^{\rm th}$ site.}
\label{fig:ivb}
\end{figure}

A heuristic picture giving the general features of $S_{imp}(r/\xi_K,r/{\mathcal R})$,
for both even and odd ${\mathcal R}$, 
can be derived from a resonating valence bond view of the ground state~\cite{Refael}. We loosely identify 
$S_{imp}$ with $\ln 2 \times$ the probability of an ``impurity valence bond'' (IVB) stretching 
from the impurity site into region $B$. (See Fig.~\ref{fig:ivb}.) For $J_K^{\prime}=0$ and even ${\mathcal R}$, 
all lengths for the IVB are equally probable so the naive picture predicts that $S_{imp}$ 
decreases linearly from $\ln 2$ to $0$ with increasing $r/R$, in rough 
agreement with the DMRG data in Fig.~\ref{fig:sfp}.  We expect the typical 
length of the IVB to be $\xi_K$ (when $\xi_K<\mathcal{R}$), leading to the monotonic decrease of $S_{imp}$ 
with increasing $r/\xi_K$, for ${\mathcal R}$ even, and its 
vanishing as $r/\xi_K\to \infty$. For ${\mathcal R}$ odd, the 
ground state contains one unpaired spin.  When $\xi_K=\infty$ this unpaired spin 
is the impurity, implying no IVB and hence $S_{imp}=0$.  As $\xi_K$ decreases 
the probability of having an IVB increases, since the unpaired spin 
becomes more likely to be at another site, due to Kondo screening, 
but the average length of the IVB, when it is present, decreases. 
These two effects trade off to give $S_{imp}$ a maximum when $\xi_K\approx R$, in good agreement 
with Fig.~\ref{fig:Scaling1} inset, at which 
point the probability of having an IVB becomes $O(1)$ but the decreasing average size of the IVB 
starts to significantly decrease the probability of it stretching into region $B$. 

\begin{figure}[!th]
\begin{center}
\includegraphics[width=0.95\columnwidth,clip]{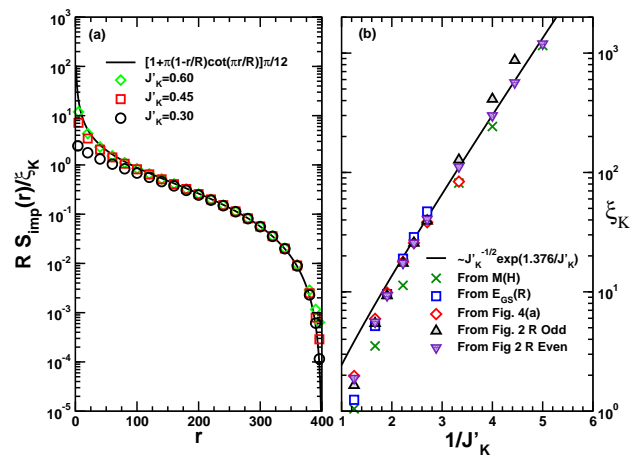}
\par
\end{center}
\caption{(a) 
DMRG results for $S_{\mathrm{imp}}(r)\mathcal{R}/\xi _{K} $ 
for the spin chain at $J^c_2$
for various values of $J_{K}^{\prime }$ in
the Fermi liquid regime. 
The black curve is the FLT prediction, Eq.~(\ref{eq:FLT_FS}).
(b) The various estimates for $\xi_K$ compared with the RG
result $\xi_K\propto\exp (1.376/J_K^\prime )/\sqrt{J_K^{\prime}}$.}
\label{fig:ScalingJ1J2}
\end{figure}

{\it FLT calculation.}
The non-interacting part of the 1$D$ (s-wave) fermion Hamiltonian can be written as the
integral over $r$ of spin and charge energy densities for left and
right-movers, $\mathcal{H}_{s,L/R}$ and $\mathcal{H}_{c,L/R}$. 
At energy
scales $\ll T_{K}$ the impurity spin is screened and does not appear in the
low energy effective Hamiltonian. Irrelevant local interactions, at $r=0$,
not involving the impurity spin appear instead \cite{IanCFT}. The most
relevant of these can be written:
\begin{equation}
H_{int}=-(\xi_K\pi )\mathcal{H}_{s,L}(0).  \label{FLT}
\end{equation} 
Various long distance, low energy properties can be calculated in an expansion 
in $\xi_K/r$ and $E\xi_K/v$ at distances $r\gg \xi_K$ and energy scales $E\ll v/\xi_K$.
For instance, the zero temperature impurity susceptibility is $\chi (T=0)=\xi_K/(4v)$.
In the limit $\xi _{K}\ll r$, we can calculate $S_{imp}$
using the FLT interaction of Eq.~(\ref{FLT}) in lowest order perturbation
theory in $\xi_K$. This calculation is easily done by CFT methods. We
first express tr$\rho ^{n}$ in terms of the partition function, $Z_{n}(r)$
on an $n$-sheeted Riemann surface, $\mathcal{R}_{n}$, with the sheets joined
at the cut extending from $r$ to $\mathcal{R}$ \cite{Wilczek94, Cardy04}. The
entanglement entropy is obtained by means of the replica trick:
\begin{equation}
S=-\lim_{n\to 1}\frac{d}{d n}[Z_{n}/Z^{n}].  \label{rep}
\end{equation}
($Z$ is the partition function on the ordinary complex plane.) The
correction to $Z_{n}$ of first order in $\xi_K$ is:
\begin{equation}
-\delta Z_{n}=-(\xi_K\pi )n\int_{-\infty }^{\infty }d\tau \langle \mathcal{H}%
_{s,L}(\tau ,0)\rangle_{\mathcal{R}_{n}}.  \label{int}
\end{equation}
$\mathcal{H}_{s,L}=T/(2\pi )$ where $T(\tau ,x)$ is the conventionally
normalized energy-momentum tensor for the $c=1$, free boson conformal field
theory corresponding to the spin excitations of the original free fermion
model. It obeys \cite{Cardy04}:
$
\langle T(\tau ,0)\rangle_{\mathcal{R}_{n}}=v\Delta _{n}(2ir)^{2}/[(v\tau
+ir)^{2}(v\tau -ir)^{2}],  \label{<T>}
$
with $\Delta _{n}=(c/24)(1-n^{-2})$. Both the integral in Eq.~(\ref{int})
and the limit in eq. (\ref{rep}) are trivial, yielding (see Fig~\ref{fig:Scaling1}):
\begin{equation}
S_{imp}=\pi \xi_K/(12r).
\label{eq:FLT_FSS}
\end{equation}
Our FLT calculation of $S_{imp}$ is
easily extended to finite $\mathcal{R}$. In this case, $\langle T\rangle$ is given 
as above
with the substitution:
$
v\tau \pm ir\to (2\mathcal{R}/\pi )\sinh [\pi (v\tau \pm ir)/(2\mathcal{R})],
$
and $2r\to (2\mathcal{R}/\pi )\sin [\pi r/(\mathcal{R})]$. The integral in
Eq.~(\ref{int}) can again be done exactly, yielding:
\begin{equation}
S_{imp}=[\pi \xi_K/(12\mathcal{R})][1+\pi (1-r/\mathcal{R})\cot (\pi r/%
\mathcal{R})].  \label{eq:FLT_FS}
\end{equation}
We note that this result can be simply obtained from $S_{U0}=(1/6)\ln                                                                        
[{\mathcal R}\sin \pi r/{\mathcal R})]$ by shifting both $r$ and $R$ by $\pi                                                                          
\xi_K/2$ and Taylor expanding to first order in the small quantities                                                                          
$\xi_K/r$ and $\xi_K/R$.
Excellent agreement between this result and DMRG data is shown 
in Fig.~\ref{fig:ScalingJ1J2}(a) allowing for a determination of $\xi_K(J'_K)$.

In Figs. \ref{fig:Scaling1},\ref{fig:ScalingJ1J2}(a) we have determined                                                                   
$\xi_K(J_K^\prime )$ by requiring good data collapse. It is an important                                                                  
consistency check on our scaling hypothesis to also determine                                                                             
$\xi_K(J_J^\prime )$ independently by more conventional methods.  Using                                                                   
the beta-function to cubic order we find that \cite{SCLA}: 
$\xi_K\propto\exp (1.376/J_K^\prime )/\sqrt{J_K^{\prime}}$. 
[To derive the $1.376$                                                                    
factor we needed the spin-wave velocity, $v\approx 1.1699$ and the                                                                        
coefficient of proportionality between $\vec S_2+J_2^c\vec S_3$ and the                                                                   
free fermion spin density at the impurity location, which was determined                                                                  
by comparing the end to end spin correlation function (decaying as                                                                        
$1/\mathcal{R}^2$)  in the spin chain and 1$D$ free fermion models.] This                                                                   
formula is only valid at very weak coupling. We also derived \cite{SCLA}                                                                  
the Fermi liquid result:
\begin{equation}
E_{\rm GS}({\mathcal R},J_K^\prime)=e_{0}+e_{1}\mathcal{R}-\frac{\pi v}{24\mathcal{R}}+\left( e_{2}+\frac{\pi
^{2}v\xi _{K}}{48}\right) \frac{1}{\mathcal{R}^{2}},
\end{equation}
(where $e_0$, $e_1$ and $e_2$ are some constants) and used                                                                 
it to extract $\xi_K(J_K^{\prime})$ from DMRG results for the ground state                                                                 
energy.  Finally, we compared the magnetization versus field of our spin                                                                  
model to exact Bethe ansatz results~\cite{Andrei} for the 1$D$ fermion model with length                                                                  
$\mathcal{R}$, obtaining excellent agreement and determining $\xi_K$                                                                      
\cite{SCLA}.  We show all our results for $\xi _{K}$ versus $1/J_K^\prime$                                                                
in Fig.~\ref{fig:ScalingJ1J2}(b) observing a good agreement between the                                                                   
different estimates and, at small $J_K^\prime$, with the weak coupling RG                                                                 
prediction.                                        

We note that, if we take region $A$ to be the impurity spin only, 
then the resulting entanglement entropy can be expressed in terms of 
the ground state impurity magnetization, $m_i$:
$
s_{imp}=-\sum_{\pm }\left( {1/2\pm m_{imp}}\right) \ln \left( {1/2\pm m_{imp}}
\right).
$
 (Here $m_i$ is only non-zero when the total number of spins is 
odd, in which case the ground state is a spin doublet.)
$m_{imp}$ was analyzed in \cite{Sorensen} for the 1$D$ fermion Kondo model and 
we find similar behavior for the spin chain model. This immediately 
determines the behavior of $s_{imp}$~\cite{SCLA}.

In conclusion we have defined an impurity entanglement entropy,
calculated it using DMRG and CFT methods and shown it to be a
universal scaling function. We have also verified that the spin chain with a
weak boundary link is indeed a realization of the Kondo model and
determined accurately the corresponding Kondo length scale ($\xi_K$) as a
function of this boundary coupling.

{\it{Acknowledgements ---}}
We are grateful to J. Cardy for interesting discussions. This
research was supported by NSERC (all authors), the CIAR (I.A.)
CFI (E.S.) and SHARCNET (E.S.).
Numerical simulations have been performed on the WestGrid network and
the SCHARCNET facility at McMaster University.

\end{document}